\documentclass[11pt]{article}

\usepackage[DIV13]{typearea}
\usepackage{amsmath} 
\usepackage{amssymb}
\usepackage{amsfonts}
\usepackage{amscd}
\usepackage{graphicx}
\usepackage{epsfig}
\usepackage[footnotesize,bf]{caption2}
\usepackage{cite}
\usepackage{bbold}
\usepackage{longtable}
\usepackage[center,footnotesize,hang]{subfigure}
\usepackage{url}

\newcommand{\eV}{\ensuremath{\:\mathrm{eV}}}

\newcommand{\GeV}{\ensuremath{\:\mathrm{GeV}}}

\newcommand{\Eqref}[1]{Eq.\eqref{#1}}

\newcommand{\Figref}[1]{Figure \ref{#1}}
\newcommand{\Tabref}[1]{Table \ref{#1}}

\newcommand{\Secref}[1]{Section \ref{#1}}
\newcommand{\Appref}[1]{Appendix \ref{#1}}

\newcommand{\Cl}[1]{\mathcal{C} _{#1}}

\newcommand{\Ord}[2]{\; ^{\circ} \mathrm{#1}_{#2}  \;}
\newcommand{\OrdCl}[1]{\; ^{\circ} \mathcal{C} _{#1} \;}
\newcommand{\Rep}[1]{\underline{\mbox{\textbf{#1}}}}
\newcommand{\MoreRep}[2]{\underline{\mbox{\textbf{#1}}} _{\mbox{\textbf{#2}}}}
\newcommand{\Groupname}[2]{$ {#1} _{#2} $}

\begin{document}

\begin{titlepage}
\begin{flushright}
SISSA 76/2008/EP
\end{flushright}
\vspace*{5mm}

\begin{center}
{\Large\sffamily\bfseries
\mathversion{bold} A Supersymmetric \Groupname{D}{4} Model for $\mu-\tau$ Symmetry
\mathversion{normal}}
\\[13mm]
{\large
A.~Adulpravitchai$^{a}$~\footnote{E-mail: \texttt{adisorn.adulpravitchai@mpi-hd.mpg.de}}, 
A.~Blum$^{a}$~\footnote{E-mail: \texttt{alexander.blum@mpi-hd.mpg.de}} and
C.~Hagedorn$^{a,b}$~\footnote{E-mail: \texttt{hagedorn@sissa.it}}} 
\\[5mm]
{\small \textit{$^a$ 
Max-Planck-Institut f\"{u}r Kernphysik\\ 
Postfach 10 39 80, 69029 Heidelberg, Germany\\
$^b$ Scuola Internazionale Superiore di Studi Avanzati (SISSA)\\
via Beirut 4, I-34014 Trieste, Italy
}}
\vspace*{1.0cm}
\end{center}
\normalsize
\begin{abstract}
\noindent We construct a supersymmeterized version of the 
model presented by Grimus and Lavoura (GL) in \cite{GL1} which predicts
$\theta_{23}$ maximal and $\theta_{13}=0$ in the lepton sector. For this purpose, we extend the
flavor group, which is $D_{4} \times Z_{2}^{(aux)}$ in the original model, to 
$D_{4} \times Z_{5}$. 
An additional difference is the absence of right-handed neutrinos.
Despite these changes the model is the same as the GL model, since $\theta_{23}$
maximal and $\theta_{13}=0$ arise through the same mismatch of $D_4$ subgroups,
$D_2$ in the charged lepton and $Z_2$ in the neutrino sector.
In our setup $D_4$ is solely broken by gauge singlets, the flavons.
We show that their vacuum structure, which leads to the prediction 
of $\theta_{13}$ and $\theta_{23}$, is a natural
result of the scalar potential. We find that the neutrino mass matrix only allows for inverted
hierarchy, if we assume a certain form of spontaneous CP violation. The quantity
$|m_{ee}|$, measured in neutrinoless double beta decay, is nearly equal to the 
lightest neutrino mass $m_3$. The Majorana phases $\phi_{1}$ and $\phi_{2}$
are restricted to a certain range for $m_3 \lesssim 0.06 \eV$.
We discuss the next-to-leading order corrections which give rise to shifts in the
vacuum expectation values of the flavons. These induce deviations from maximal atmospheric
mixing and vanishing $\theta_{13}$. It turns out that these deviations are smaller for $\theta_{23}$
than for $\theta_{13}$.
\end{abstract}

\end{titlepage}

\setcounter{footnote}{0}

\section{Introduction}
\label{sec:intro}

Experiments revealed that neutrinos have properties very distinct from the other known fermions.
Their masses are much smaller and their hierarchy is much less pronounced. Unfortunately,
only two mass squared differences, the solar and the atmospheric one, are known from experiments
\cite{experimentaldata}
\begin{equation}
\Delta m_{21}^{2} = \left( 7.65 ^{+0.46} _{-0.40} \right) \cdot 10^{-5} \, \eV^2   
\;\;\; \mbox{and} \;\;\; 
|\Delta m_{31}^{2}| = \left( 2.40 ^{+0.24} _{-0.22} \right) \cdot 10^{-3} \, \eV^2
\;\;\;\;\;\;\;\; (2 \, \sigma)
\end{equation}
where $\Delta m_{ij}^{2}$ denotes $m_i^2 - m_j^2$ with $m_{i,j}$ being the neutrino masses. As the sign
of $\Delta m_{31}^{2}$ is unknown it is not clear whether neutrinos follow a normal or an inverted 
hierarchy. Even more surprising than the neutrino masses is the fact that
leptons have a very peculiar mixing pattern
with two large mixing angles and one small one. The experimental measurements
\cite{experimentaldata}
\begin{eqnarray}\nonumber
&&\sin^2\theta_{13}\leq 0.040 \;\; , \;\;\;\;\;\;\;\;\;\;\;\;\;\;\, \theta_{13}\leq 11.5^\circ
 \;\; , \;\;\;\;\;\;\;\;\;\;\;\;\;\;\;\;\;\;\;\; \theta_{13}\leq 0.2 \;\; ,
\\ 
&&\sin^2\theta_{23}=0.50^{+0.13}_{-0.11} \;\; , \;\;\;\;\;\;\;\;\; \theta_{23}=(45.0^{+7.5}_{-6.4})^\circ 
 \;\; , \;\;\;\;\;\;\;\;\;\;\; \theta_{23}= 0.785 ^{+0.13}_{-0.11} \;\; ,
\;\;\;\;\;\;\;\;\;\;\;\; (2 \, \sigma)
\\ \nonumber
&&\sin^2\theta_{12}=0.304^{+0.046}_{-0.034} \;\; , \;\;\;\;\;\; \theta_{12}=(33.5^{+2.8}_{-2.2})^\circ
 \;\; , \;\;\;\;\;\;\;\;\;\;\; \theta_{12}= 0.58 ^{+0.05}_{-0.04} \;\; ,
\end{eqnarray}
allow for the possibility that the atmospheric mixing angle $\theta_{23}$ is maximal and
the reactor mixing angle $\theta_{13}$ vanishes. These values can be deduced from a neutrino
mass matrix which is $\mu-\tau$ symmetric \cite{mutau}, i.e. does not change its form if second and third
columns and rows are interchanged, in the charged lepton mass basis.  
At the same time the solar mixing angle $\theta_{12}$ is left undetermined.
An even more constraining pattern is the one of tri-bimaximal (TB) mixing \cite{hps} in which, apart from
$\theta_{23}= \pi/4$ and $\theta_{13}= 0$, also $\theta_{12}$ is fixed by
$\sin ^{2} \theta_{12}=1/3$. Both patterns have been subject to
extensive studies in order to find a theoretical explanation. The most promising one
seems to be the assumption of an additional flavor symmetry which is responsible for such a mixing
pattern. For TB mixing the simplest models are based on the tetrahedral
group $A_{4}$ \cite{A4}, 
\footnote{From a group theoretical point of view $S_4$, the permutation group of four distinct objects,
might be even more appropriate \cite{LamS4}.}
while for $\mu-\tau$ symmetry even smaller groups are appropriate such as
the dihedral groups $D_{3} \cong S_3$ \cite{GLS3} and $D_{4}$ \cite{GL1}. 

An interesting observation which
has been made first in the $A_4$ models and then also in the models predicting $\mu-\tau$ symmetry
with a dihedral flavor group is the fact that the VEVs 
of a certain subset of scalar fields $\{ \phi_{e} \}$, coupling only to charged leptons at leading order (LO),
preserve one subgroup of the original symmetry, while another set of scalars $\{ \phi_{\nu} \}$, 
coupling only to neutrinos at LO, breaks the flavor group to a different subgroup. This mismatch 
can be regarded as an intuitive explanation for sizable mixings in the lepton sector.
\footnote{A similar observation has been made concerning the Cabibbo angle in the quark sector
whose value might be the result of a non-trivial breaking of a dihedral group such as $D_7$ or
$D_{14}$ \cite{thetaC,symmetrygeneral,dntheory}.} \footnote{Notice that there are also models which can
predict lepton mixings without preserving non-trivial subgroups
in all sectors of the theory, for instance \cite{S3alt}.} For more general considerations
on the origin of a certain mixing pattern see \cite{symmetrygeneral}.

One of the main
issues in these models is the vacuum alignment of the flavor symmetry breaking fields,
since without a special alignment the mixing pattern is merely a result of parameter tuning.
In a class of $A_4$ models \cite{A4alignment} this alignment is achieved in a supersymmetric 
framework where the scalars transforming non-trivially under flavor are
only gauge singlets. Their vacuum expectation 
values (VEVs) are driven by another set of gauge singlets, the driving fields. 
This mechanism to align the flavon VEVs is also used in models with other discrete 
\cite{otherdisc} and with continuous flavor symmetries \cite{contsymm}.

The $D_4$ model \cite{GL1} by Grimus and Lavoura (GL), which successfully predicts $\mu-\tau$ symmetry
in a natural way including a profound explanation for the vacuum alignment, is non-supersymmetric
in its original version. For several reasons, it would be desirable to supersymmeterize
this model. In doing this, it is of advantage to break the flavor and the gauge group (spontaneously)
by separate sets of scalar fields, flavons and Higgs doublets. 
In this paper we present such a supersymmeterized version  
in which the vacuum alignment is
achieved in a similar way as in \cite{A4alignment}. We arrive at $\theta_{23}=\pi/4$ and $\theta_{13}=0$
at LO and analyze the next-to-leading order (NLO) effects which will perturb this result in a particular way.
In the minimal supersymmetric extension presented here the flavor symmetry $D_4$ 
is accompanied by a $Z_5$ symmetry which plays a similar role as $Z_2^{(aux)}$  
in the original model. Furthermore, our model does not incorporate right-handed neutrinos so that 
the light neutrino masses stem from the 
dimension-5 operator $l h_u l h_u/\Lambda$.
Despite these changes the model is essentially a supersymmeterized version of the GL model,
since the prediction of maximal $\theta_{23}$ and vanishing $\theta_{13}$ is still due to the 
fact that we preserve, at LO, a $D_2$ subgroup of $D_4$ in the charged lepton 
and a $Z_2$ subgroup in the neutrino sector. Since this $Z_2$ group
 is not contained in the $D_2$ group of the charged lepton sector, $D_4$ is completely 
broken in the whole theory. Apart from predicting the value of 
$\theta_{13}$ and $\theta_{23}$, the original GL model also predicts that neutrinos
are normally ordered and that the effective Majorana mass of neutrinoless $\beta\beta$ ($0\nu\beta\beta$)
decay is equal to $|m_{ee}|= m_1 m_2/m_3$.
These predictions result from the fact that in the GL model not all fields are present which are
allowed to have a non-vanishing VEV in accordance with a preserved $Z_2$ subgroup in the neutrino sector.
Here we will include all flavons allowed by the symmetry principle
so that we still predict $\theta_{23}= \pi/4$
and $\theta_{13}=0$, but now can accommodate both mass hierarchies. 
In this respect our results are analogous to another model by GL predicting
$\theta_{13}$ and $\theta_{23}$ with the help of the dihedral group $D_{3} \cong S_3$ \cite{GLS3}.

In our phenomenological study we concentrate on the possibility of a certain type of spontaneous CP
violation in order to make the model more predictive. We find that
neutrinos then have to have an inverted hierarchy and $|m_{ee}| \approx m_3$ holds. 
Furthermore, the Majorana phases $\phi_{1,2}$ can only take values in a limited range for $m_3 \lesssim 0.06 \eV$. If we additionally
remove one of the flavons from the model (this is analogous to what is done in \cite{GL1}), the three
parameters of the model in the neutrino sector are determined by the three measured quantities,
$\Delta m_{21}^{2}$, $|\Delta m_{31}^{2}|$ and $\theta_{12}$, and the Majorana phases are predicted to be $\phi_1=\pi/2$ and $\phi_2=0$. 
The NLO corrections coming from the inclusion of operators with one more flavon
lead to deviations from $\theta_{23}=\pi/4$ and $\theta_{13}=0$. It turns out that 
$\theta_{23} -\pi/4$ is much smaller than $\theta_{13}$.

In \cite{Ishimori_D4} it has already been attempted to build a $D_4$ model in which the Higgs doublets
transforming non-trivially under the flavor group are replaced by flavons. Since this model
is non-supersymmetric the vacuum alignment problem is not straightforward to solve 
and indeed one has to require that one of the quartic couplings in the potential vanishes. However, such
an assumption will not be stable against corrections and has to be considered as a severe tuning.
In a second version of this model \cite{Ishimori_D4_SUSY} which is supersymmetric, the potential is
not studied such that the question of the vacuum alignment also remains open. Hence, a successful
supersymmeterization of the original $D_4$ model by GL still does not exist. 

The paper is organized as follows: in \Secref{sec:grouptheory} we repeat the necessary group theory
of $D_4$ and the properties of the subgroups relevant in the $D_4$ model. \Secref{sec:LO} contains 
the LO results for the lepton masses and mixings as well as the 
flavon potential. In the following section the NLO corrections are studied.
We summarize our results and give a short outlook in
\Secref{sec:summary}. In the two Appendices we treat additional group theoretical
aspects of the model.

\mathversion{bold}
\section{Group Theory of $D_4$}
\mathversion{normal}
\label{sec:grouptheory}

In this section we briefly review basic features of the dihedral group $D_4$.
Its order is eight, and it has five irreducible representations which we denote as
$\MoreRep{1}{i}$, $\rm i=1,...,4$ 
and $\Rep{2}$. All of them are real and only $\Rep{2}$
is faithful. The group is generated by the two generators $\rm A$ and $\rm B$
which can be chosen as \cite{dngrouptheory}
\begin{equation}
\label{eq:generators}
\rm A =\left(\begin{array}{cc} 
                            i & 0 \\
                            0 & -i 
          \end{array}\right) \;\;\; \mbox{and} \;\;\; \rm 
		B=\left(\begin{array}{cc} 
                                       0 & 1 \\
                                       1 & 0 
                  \end{array}\right) 
\end{equation}
for $\Rep{2}$. Note that $\rm A$ is a complex matrix, although $\Rep{2}$ is a real
representation. For $(a_1,a_2)^{T} \sim \Rep{2}$ therefore $(a_2^{\star}, a_{1}^{\star})^{T}$
transforms as $\Rep{2}$ under $D_4$. The generators of the one-dimensional
representations can be found in the character table, displayed in \Tabref{tab:chartab}.
The generators fulfill the relations
\begin{equation}\label{eq:genrelations}
\mathrm{A}^{4} =\mathbb{1} \;\;\; , \;\;\; \rm B^2=\mathbb{1} \;\;\; \mbox{and}
\;\;\; \rm ABA=B \; . 
\end{equation}
\begin{table}
\begin{center}
\begin{tabular}{|l|ccccc|}
\hline
&\multicolumn{5}{|c|}{classes}\\ 
\cline{2-6}
&$\Cl{1}$&$\Cl{2}$&$\Cl{3}$&$\Cl{4}$&$\Cl{5}$\\
\hline
\rule[0.1cm]{0cm}{0cm} $\rm G$                 &$\rm \mathbb{1}$&$\rm
A$&$\rm A^{2}$ & $\rm B$ & $\rm A \, B$  \\
\hline
$\OrdCl{i}$          &1      &2   &1   &2  &2 \\
\hline
$\Ord{h}{\Cl{i}}$    &1      &4      &2 & 2 & 2 \\
\hline
\rule[0in]{0.3cm}{0cm}$\MoreRep{1}{1}$    &1 & 1 & 1 & 1 & 1 \\                             
\rule[0in]{0.3cm}{0cm}$\MoreRep{1}{2}$    &1 & 1 & 1 & -1 & -1 \\   
\rule[0in]{0.3cm}{0cm}$\MoreRep{1}{3}$    &1 & -1 & 1 & 1 & -1\\                             
\rule[0in]{0.3cm}{0cm}$\MoreRep{1}{4}$    &1 & -1 & 1 & -1 & 1\\                            
\rule[0in]{0.3cm}{0cm}$\Rep{2}$    	  &2 & 0 & -2 & 0 & 0\\[0.1cm]
\hline
\end{tabular}
\end{center}
\begin{center}
\normalsize
\begin{minipage}[t]{12cm}
\caption[Character table of the group
  \Groupname{D}{4}]{Character table of the group
  \Groupname{D}{4}. $\Cl{i}$ are the classes of the
group, $\OrdCl{i}$ is the order of the $i ^{\mathrm{th}}$ class, i.e. the number of distinct elements contained in this class, $\Ord{h}{\Cl{i}}$
is the order of the elements $S$ in the class $\Cl{i}$, i.e. the smallest
integer ($>0$) for which the equation $S ^{\Ord{h}{\Cl{i}}}= \mathbb{1}$
holds. Furthermore the table contains one representative for each
class $\Cl{i}$ given as product of the generators $\rm
A$ and $\rm B$ of the group. \label{tab:chartab}}
\end{minipage}
\end{center}
\end{table}
The product rules for $\MoreRep{1}{i}$ are the following
\begin{equation}\nonumber
\MoreRep{1}{i} \times \MoreRep{1}{i}= \MoreRep{1}{1} \; , \;\; 
\MoreRep{1}{1} \times \MoreRep{1}{i}= \MoreRep{1}{i} \;\; \mbox{for} \;\; \rm i=1,...,4 
\; , \;\;
\MoreRep{1}{2} \times \MoreRep{1}{3}= \MoreRep{1}{4} \; , \;\;
\MoreRep{1}{2} \times \MoreRep{1}{4}= \MoreRep{1}{3} \;\; \mbox{and} \;\;
\MoreRep{1}{3} \times \MoreRep{1}{4}= \MoreRep{1}{2} \; .
\end{equation}
For $s_i \sim \MoreRep{1}{i}$ and $(a_1,a_2)^{T} \sim \Rep{2}$ we find
\begin{equation}\nonumber
\left( \begin{array}{c} s_1 a_1 \\ s_1 a_2
\end{array} \right) \sim \Rep{2} \;\; , \;\;\;
\left( \begin{array}{c} s_2 a_1 \\ -s_2 a_2
\end{array} \right) \sim \Rep{2} \;\; , \;\;\;
\left( \begin{array}{c} s_3 a_2 \\ s_3 a_1
\end{array} \right) \sim \Rep{2} \;\;\; \mbox{and} \;\;\;
\left( \begin{array}{c} s_4 a_2 \\ -s_4 a_1
\end{array} \right) \sim \Rep{2} \;\; .
\end{equation}
The product $\Rep{2} \times \Rep{2}$ decomposes into the four singlets
which read for $(a_1,a_2)^{T}$, $(b_1, b_2)^{T}$ $\sim \Rep{2}$
\begin{equation}\nonumber
a_1 b_2 + a_2 b_1 \sim \MoreRep{1}{1} \;\; , \;\;\;
a_1 b_2 - a_2 b_1 \sim \MoreRep{1}{2} \;\; , \;\;\;
a_1 b_1 + a_2 b_2 \sim \MoreRep{1}{3} \;\;\; \mbox{and} \;\;\;
a_1 b_1 - a_2 b_2 \sim \MoreRep{1}{4} \;\; .
\end{equation}
More general formulae for generators, Kronecker products and Clebsch Gordan
coefficients can be found, for example, in \cite{dntheory,kronprods}.
Notice that our group basis does not coincide with the one chosen by GL in \cite{GL1}.
Therefore, the mass matrices shown below have another appearance, especially the charged
lepton mass matrix is not diagonal in our basis. However, the
prediction of the mixing angles does not depend on the chosen group basis.
In \Appref{appA} we explicitly discuss the correlation between our basis and the one found in
\cite{GL1}.

All subgroups of $D_4$ are abelian: $Z_2 \cong D_1$, $Z_4$ and $D_2 \cong Z_2 \times Z_2$.
We are interested here in $Z_2$ subgroups which are generated by $\mathrm{B} \, \mathrm{A}^{m}$
with $m=0,...,3$ and the $D_2$ subgroup generated by $\rm A^{2}$ and $\rm B A$. 
In order to see that $\mathrm{B} \, \mathrm{A}^{m}$ gives a $Z_2$ group note that
\begin{equation}\nonumber
(\mathrm{B} \, \mathrm{A}^{m})^{2} = \mathrm{B} \, \mathrm{A}^{m} \mathrm{B} \, \mathrm{A}^{m}
= \mathrm{B} \, \mathrm{A}^{m-1} \mathrm{B} \, \mathrm{A}^{m-1} = \dots 
= \mathrm{B}^{2} = \mathbb{1}
\end{equation}
holds, if \Eqref{eq:genrelations} is used. Similarly, one finds for 
$\rm A^{2}$ and $\rm B A$ 
\begin{equation}\nonumber
(\rm A^{2})^{2} = A^{4} = \mathbb{1} \;\;\; \mbox{and} \;\;\;  (B A)^{2} = B A B A = B^2 = \mathbb{1}
\end{equation}
by using again the generator relations. Obviously, $\rm A^2$ and $\rm B A$ are not equal (in general) and 
thus they generate different $Z_2$ subgroups.
Additionally, we have to check that $\rm A^{2}$ and $\rm B A$ commute
\begin{equation}\nonumber
\rm A^{2} B A = A^{3} B A^{2} = A^{4} B A^{3} = B A A^{2}  \;\; .
\end{equation}
All this shows that $\rm A^2$ and $\rm B A$ generate a $Z_2 \times Z_2$ group
which is isomorphic to a $D_2$ group.
The other non-trivial
element of the $D_2$ group is $\rm B A^{3}$. Thus, one could also use the two elements
$\rm A^2$ and $\rm B A^3$ to generate this $D_2$ group.
However, we follow the convention to use as generators $\rm A^2$ and the element $\mathrm{B} \mathrm{A}^{p}$
with $p$ being the smallest possible natural number.
The $Z_2$ symmetry given through $\mathrm{B} \mathrm{A}^{m}$ is
left unbroken by a non-vanishing VEV of a singlet transforming as $\MoreRep{1}{3}$ if $m$ is 
even and of one transforming as $\MoreRep{1}{4}$ for $m$ being odd. 
Additionally, it is left intact by fields $\psi_{1,2}$ 
forming a doublet, if their VEVs have the following structure
\begin{equation}
\left( \begin{array}{c} \langle \psi_{1} \rangle \\
 \langle \psi_{2} \rangle \end{array} \right) \propto \left(
\begin{array}{c} \mathrm{e}^{-\frac{\pi i m}{2}}\\ 1
\end{array}
\right) \; .
\end{equation}
For preserving the $D_2$ group generated by $\rm A^2$ and $\rm B A$ only singlets in 
$\MoreRep{1}{4}$ are allowed to have a non-vanishing VEV. 
Especially, no fields forming a doublet under $D_4$ should acquire a VEV.
Clearly, in all cases
singlets in the trivial representation of $D_4$, $\MoreRep{1}{1}$, are allowed to 
have a non-vanishing VEV. Note also that in none of the cases a field transforming as $\MoreRep{1}{2}$
can acquire a non-zero VEV. Since we concentrate on the $D_2$ subgroup induced by $\rm A^2$
and $\rm B A$, the $Z_2$ subgroup has to be generated by $\mathrm{B} \mathrm{A}^m$ with $m$
being even in order not to be a subgroup of the $D_2$ group. Only then the mismatch between the
two subgroups is achieved. The choice of $m$, $m=0$ or $m=2$, depends 
on the relative sign between $\langle\psi_{1} \rangle$ and $\langle\psi_{2} \rangle$ 
for two fields $\psi_{1,2} \sim \Rep{2}$.

\section{The Model at Leading Order}
\label{sec:LO}

We augment the Minimal Supersymmetric Standard Model (MSSM) by the flavor symmetry $D_4 \times Z_5$. 
As mentioned above, the non-trivial breaking of
$D_4$ is responsible for maximal atmospheric mixing and vanishing $\theta_{13}$, while $Z_5$ is necessary to separate 
the charged lepton and the neutrino sector. The model contains
three left-handed lepton doublets $l_i$, the three right-handed charged leptons $e^c_i$,
the MSSM Higgs doublets $h_{u,d}$
and two sets of flavons $\{ \chi_e, \varphi_e\}$ and $\{ \chi_\nu, \varphi_\nu, \psi_{1,2}\}$ which break $D_4$
in the charged lepton and the neutrino sector, respectively. The transformation properties of these fields
are collected in \Tabref{tab:particles}.
\begin{table}
\begin{center}
\begin{tabular}{|c||c|c|c|c||c|c||c|c|c|c|c|c|c|}\hline
Field & $l_{1}$ & $l_{2,3}$ & $e^{c}_{1}$ & 
$e^{c}_{2,3}$ & $h_{u}$ & $h_{d}$
& $\chi_{e}$ & $\varphi_{e}$ & $\chi_{\nu}$ & $\varphi_{\nu}$
& $\psi_{1,2}$ \\ 
\hline
$D_4$ & $\MoreRep{1}{1}$ & 
$\Rep{2}$ & $\MoreRep{1}{1}$ & $\Rep{2}$ & $\MoreRep{1}{1}$ & $\MoreRep{1}{1}$ 
 & $\MoreRep{1}{1}$ & $\MoreRep{1}{4}$ & $\MoreRep{1}{1}$& $\MoreRep{1}{3}$
& $\Rep{2}$ \\
$Z_5$ & $\omega$ & $\omega$ & 
$1$ & $1$ & $\omega^{3}$ & $\omega$ & $\omega^3$ & $\omega^3$
& $\omega^2$ & $\omega^2$ & $\omega^2$\\ 
\hline
\end{tabular}
\end{center}
\begin{center}
\begin{minipage}[t]{12cm}
\caption[]{Particle content of the model. $l_i$ denotes the three left-handed lepton $SU(2)_L$ doublets,
$e^c_i$ are the right-handed charged leptons and $h_{u,d}$ are the MSSM Higgs doublets.
The flavons $\chi_e$, $\varphi_e$, $\chi_\nu$, $\varphi_\nu$ and $\psi_{1,2}$ only transform
under $D_4 \times Z_5$. The phase factor $\omega$ is $e^{\frac{2 \pi i}{5}}$.
\label{tab:particles}}
\end{minipage}
\end{center}
\end{table}

\subsection{Fermion Masses}
\label{sec:fermionmasses_LO}

The invariance of the charged lepton and neutrino mass terms under the flavor group $D_4 \times Z_5$
requires the presence of at least one flavon. Thus, charged lepton masses are generated 
by non-renormalizable operators only. In a model which treats quarks as well this allows the explanation 
of the small $\tau$ mass compared to the top quark mass without relying on a large value of 
$\tan \beta = \langle h_u \rangle/\langle h_d \rangle= v_u/v_d$. 
The neutrinos receive Majorana masses through the dimension-5 operator $l h_u l h_u/\Lambda$
which can be made invariant under the flavor group by coupling to a flavon. 
The part of the superpotential giving lepton masses reads at LO
\begin{eqnarray}
\label{eq:wlatLO}
w_l &=& y_1^e \chi_{e} l_1 e_1^c \frac{h_d}{\Lambda}
       + y_2^e \chi_e (l_2 e_3^c + l_3 e_2^c) \frac{h_d}{\Lambda}
       + y_3^e \varphi_e (l_2 e_2^c - l_3 e_3^c) \frac{h_d}{\Lambda}\\
    & & + y_1 \chi_{\nu} l_1 l_1 \frac{h_u^2}{\Lambda^2}
        + y_2 l_1 (l_2 \psi_2 + l_3 \psi_1) \frac{h_u^2}{\Lambda^2}
	+ y_2 (l_2 \psi_2 + l_3 \psi_1) l_1 \frac{h_u^2}{\Lambda^2}
        + y_3 \varphi_{\nu} (l_2 l_2 + l_3 l_3) \frac{h_u^2}{\Lambda^2}\nonumber \\ 
    & & + y_4 \chi_{\nu} (l_2 l_3 + l_3 l_2) \frac{h_u^2}{\Lambda^2}. \nonumber 
\end{eqnarray}
$\Lambda$ is the cutoff scale of the theory whose order of magnitude is determined by the scale of the
light neutrino masses, see below. For the moment we assume that the flavons $\chi_e$ and $\varphi_e$ 
acquire the VEVs
\begin{equation}
\label{eq:VEVs1}
\langle \varphi_e \rangle = u_e \;\;\; \mbox{and} \;\;\;
\langle \chi_e \rangle = w_e \; .
\end{equation}
As discussed in \Secref{sec:grouptheory} these VEVs break $D_4$ down to $D_2$ generated by
$\rm A^2$ and $\rm B A$ in the charged lepton sector.
The VEVs of the flavons coupling only to neutrinos at LO, are of the form
\begin{equation}
\label{eq:VEVs2}
\langle \varphi_{\nu} \rangle = u \; , \;\; \langle \chi_{\nu} \rangle = w \; , \;\;
\left( \begin{array}{c} \langle \psi_1 \rangle \\ \langle \psi_2 \rangle \end{array} \right) 
=  v \, \left( \begin{array}{c} 1 \\ 1 \end{array} \right),
\end{equation}
and therefore leave a $Z_2$ subgroup, generated by $\rm B$, unbroken. As mentioned, the equality of
the VEVs of $\langle \psi_1 \rangle$ and $\langle \psi_2 \rangle$ is crucial. 
As will be discussed in \Secref{sec:flavons_LO},
the vacuum structure in \Eqref{eq:VEVs1} and \Eqref{eq:VEVs2} is a natural result of the minimization of the 
flavon potential. We obtain the following fermion mass matrices, when inserting the flavon VEVs and $\langle h_{u,d}
\rangle = v_{u,d}$ 
\begin{equation}
\label{eq:fermionsatLO}
M_{l} = \frac{v_d}{\Lambda} \, 
\left( 
\begin{array}{ccc}  
y_1^e w_e & 0 & 0 \\ 
0  & y_3^e u_e &  y_2^e w_e \\ 
0 &  y_2^e w_e & -y_3^e u_e 
\end{array} 
\right) \;\;\; \mbox{and} \;\;\;
M_{\nu} = \frac{v^2_u}{\Lambda^2} \, 
\left( 
\begin{array}{ccc}  
y_1 w & y_2 v & y_2 v \\ 
y_2 v & y_3 u &  y_4 w \\ 
y_2 v &  y_4 w & y_3 u 
\end{array} 
\right) \; .
\end{equation}
Thereby, the left-handed fields are on the left-hand and the right-handed fields on the right-hand side for $M_l$.
The matrix $M_l \,M_l^{\dagger}$ is diagonalized through the unitary matrix $U_l$, i.e.
$U_l^{\dagger} \, M_l \, M_l^{\dagger} \, U_l$ is diagonal. $U_l$ acts on the left-handed charged lepton fields 
and is given by
\begin{equation}
\label{eq:Ul}
U_l = 
\left( 
\begin{array}{ccc} 
1 & 0 & 0 \\ 
0 & e^{i \pi /4}/\sqrt{2} & e^{-i \pi /4}/\sqrt{2} \\ 
0  & e^{-i \pi /4}/\sqrt{2}  & e^{i \pi /4}/\sqrt{2} 
\end{array} 
\right) \; .
\end{equation}
For the masses of the charged leptons we find
\begin{equation}
\label{eq:leptonmassesLO}
m_e = \frac{v_d}{\Lambda} \vert  y_1^e w_e \vert \; ,
m_{\mu} = \frac{v_d}{\Lambda} \vert  y_3^e u_e+iy_2^e w_e \vert \;\;\; \mbox{and} \;\;\;
m_{\tau} =\frac{v_d}{\Lambda} \vert  y_3^e u_e-iy_2^e w_e \vert \; .
\end{equation}
In order to arrive at non-degenerate masses for the $\mu$ and the $\tau$ lepton either  $y_3^e u_e$ or
 $y_2^e w_e$ 
has to be non-real indicating CP violation in the Yukawa couplings and/or flavon VEVs.
For $m_\tau$ being around $2 \GeV$ we find that
for small $\tan \beta$ - corresponding to $v_d$ of the order of $100 \GeV$ - the ratio of the flavon VEVs $u_e$
and $w_e$ over the cutoff scale $\Lambda$ should fulfill 
\footnote{Although not excluded, there is no obvious reason to assume that there is a large  
hierarchy among the different flavon VEVs. In general, these are correlated through the parameters
of the flavon potential.}
\begin{equation}
\label{eq:VEVestimate}
\frac{u_e}{\Lambda} \, , \frac{w_e}{\Lambda} \sim \lambda^2 \approx 0.04
\end{equation}
with $\lambda$ being the Cabibbo angle. The smallness of the ratio $m_e/m_\tau$ 
is in this model only 
explained by the assumption of a small enough coupling $y_1^e$. Similarly,
$m_\mu/m_\tau$  enforces a certain cancellation between the
two contributions $y_3^e u_e$ and $i y_2^e w_e$ in $m_\mu$. 
In \cite{GL1} these problems have been solved
by the assumption that the electron couples to a Higgs field different from those coupling to the
$\mu$ and the $\tau$ lepton and by an additional symmetry which leads to $m_\mu=0$, if it is 
unbroken, respectively.

The neutrino mass matrix in the charged lepton mass basis reads
(indicated by a prime ($\; ^\prime\;$))
\begin{equation}
\label{eq:Mnu}
M_{\nu}'= U_l^{\dagger} M_{\nu} U_{l}^{\ast} 
= \frac{v^2_u}{\Lambda^2} 
\left( 
\begin{array}{ccc} 
y_1 w & y_2 v  & y_2 v \\ 
y_2 v & y_4 w & y_3 u \\ 
y_2 v  & y_3 u & y_4 w 
\end{array} 
\right).
\end{equation}
As $M_\nu^\prime$ is $\mu-\tau$ symmetric, it immediately follows that the lepton mixing angle $\theta_{13}$
vanishes and $\theta_{23}$ is maximal. The solar mixing angle $\theta_{12}$ is not predicted, 
but in general expected to be large. Also the Majorana phases $\phi_{1,2}$ are not constrained.
The lepton mixing matrix is of the form
\begin{equation}
U_{MNS}= 
\mathrm{diag}(e^{i \gamma_1},e^{i \gamma_2}, e^{i \gamma_3})
\cdot \left( 
\begin{array}{ccc} 
\cos \theta_{12} & \sin \theta_{12} & 0  \\ 
-\frac{\sin \theta_{12}}{\sqrt{2}} & \frac{\cos \theta_{12}}{\sqrt{2}} 
& -\frac{1}{\sqrt{2}}\\ 
-\frac{\sin \theta_{12}}{\sqrt{2}} & \frac{\cos \theta_{12}}{\sqrt{2}} 
& \frac{1}{\sqrt{2}} 
\end{array} \right) \cdot \mathrm{diag}(e^{i \beta_1},e^{i \beta_2}, e^{i \beta_3}) \; .
\end{equation}
The Majorana phases $\phi_{1,2}=\alpha_{1,2}/2$
can be extracted from $U_{MNS}$ by bringing it into the standard form \cite{pdg}. 
Due to the additional factor $1/2$ the phases $\phi_{1,2}$ vary between $0$ and $\pi$. 
Assuming that all flavon VEVs are of the same size, the estimate in
\Eqref{eq:VEVestimate} also holds for the VEVs of the flavons $\chi_\nu$, $\varphi_\nu$ and $\psi_{1,2}$. 
For small $\tan \beta$, i.e. $v_u \approx v_d \approx 100 \GeV$, 
a light neutrino mass scale between $\sqrt{|\Delta m_{31}^{2}|} \approx 0.05 \eV$ and $1 \eV$ 
fixes the range of the cutoff scale $\Lambda$ to be  
\begin{equation}
4 \cdot 10^{11} \GeV \lesssim \Lambda \lesssim 8 \cdot 10^{12} \GeV \; .
\end{equation}

As shown in \Secref{sec:flavons_LO}, 
we can assume that CP is only spontaneously violated in this model by imaginary VEVs $w_e$ and $w$ of 
$\chi_e$ and $\chi_\nu$. Thus,
apart from $w_e$ and $w$ all other parameters, i.e. couplings and VEVs, are real in the following.
According to \Eqref{eq:leptonmassesLO} an imaginary $w_e$ allows the $\mu$ and the $\tau$ lepton mass to be non-degenerate.  
In the neutrino sector only the VEV $w$
of $\chi_\nu$ is imaginary, whereas all other entries in $M_\nu^\prime$ are real, so that the
matrix in \Eqref{eq:Mnu} can be written as
\begin{equation}
\label{eq:Mnu_SCPV}
M_{\nu}'= \frac{v^2_u}{\Lambda} \, \frac{v}{\Lambda} \,
\left( 
\begin{array}{ccc} 
i \, s & t & t \\ 
t & i \, x & z \\ 
t  & z & i \, x 
\end{array} 
\right)
\end{equation}
where we define the real parameters
\begin{equation}
\label{eq:Mnuparameters}
s = y_1 \frac{\mathrm{Im} (w)}{v} \; , \;\;
t = y_2 \; , \;\;
x = y_4 \frac{\mathrm{Im} (w)}{v} \;\;\; \mbox{and} \;\;\;
z = y_3 \frac{u}{v} \; .
\end{equation}

\subsection{Phenomenology}
\label{sec:phenomenology_LO}

In the following we analyze the phenomenology of this model. 
For the eigenvalues of $M_\nu' M_{\nu}'^\dagger$ we find
\begin{eqnarray}
\label{eq:numasses} \nonumber
&m_{2,1}^2 = \frac{1}{2} \, \Big( \frac{v_u^2}{\Lambda} \Big)^2 \, \Big( \frac{v}{\Lambda} \Big)^{2} \, 
\Big[ s^2+4t^2+x^2+z^2 \pm\sqrt{(s-x)^2 (8t^2+(s+x)^2)+2 (4 t^2+x^2-s^2)z^2+z^4} \Big]& \\
&\;\;\; \mbox{and} \;\;\; m_3^2 =  \Big( \frac{v_u^2}{\Lambda} \Big)^2 \, \Big( \frac{v}{\Lambda} \Big)^2 \,  \Big( x^2 +z^2 
\Big) \; .&
\end{eqnarray}
This assignment of the eigenvalues is unambiguous, since 
$m_2^2 > m_1^2$ is experimentally known and the eigenvalue corresponding to the eigenvector
$(0,1,-1)^{T}$ can only be $m_3^2$.
The solar mixing angle $\theta_{12}$ is found to depend on $s$, $t$, $x$ and $z$ in the following way
\begin{equation}
\label{eq:tanth12}
\tan{2 \theta_{12}} = \frac{2 \sqrt{2} \, |t| \, \sqrt{(s-x)^2+z^2}}{x^2 + z^2 -s^2} \; .
\end{equation}

Before discussing the general case with unconstrained parameters $s$, $t$, $x$ and $z$ we comment
on the special case in which $z$ vanishes, since then the model contains three real parameters which 
can be determined by the three experimental quantities $\Delta m_{21}^2$, $|\Delta m_{31}^2|$ and
$\theta_{12}$. According to \Eqref{eq:Mnuparameters} either $y_3$ or $u$ have to vanish for $z=0$
to hold. Assuming that $y_3$ is zero however has to be regarded as fine-tuning. 
In contrast to that, a vanishing VEV $u$ can  
be explained either through the absence of the flavon $\varphi_{\nu}$ from the model or
through a flavon potential which only allows configurations with $u=0$ to be minima.
The neutrino mass $m_3$ is then proportional to $|x|$.
From \Eqref{eq:numasses} and \Eqref{eq:tanth12} we can derive for $z=0$ 
\begin{equation}
\label{eq:m3tanth12rel_2}
m_3^2 = - \frac{1}{4} \, \frac{\cos^4 \theta_{12}}{\sin^2 \theta_{12}} \, 
\frac{(\Delta m_{21}^2 + \Delta m_{31}^2 (\tan^4{\theta_{12}}-1))^2}{\Delta m_{31}^2 (1+\tan^2{\theta_{12}})
-\Delta m_{21}^2} \; .
\end{equation}
Neglecting the solar mass squared difference we can simplify this expression to
\begin{equation}
\label{eq:m3tanth12rel}
m_3^2 \approx  - \Delta m_{31}^2 \cot^2{2 \theta_{12}} \; .
\end{equation}
\Eqref{eq:m3tanth12rel} shows that $\Delta m_{31}^2<0$, i.e. 
the neutrinos have to have an inverted hierarchy. 
Note that similar results can also be found in \cite{Merle:2006du}.
A relation analogous to \Eqref{eq:m3tanth12rel_2} can be found for 
$|m_{ee}|$ measured in $0\nu\beta\beta$ decay experiments. Note
that $|m_{ee}|$ is proportional to $|s|$ due to \Eqref{eq:Mnu_SCPV} and can be written in terms of $m_3$, $\tan \theta_{12}$ and the mass squared
differences as
\begin{equation}
|m_{ee}|^2 = m_3^2 \frac{(\Delta m_{21}^2 (1-2 \tan^2{\theta_{12}})+\Delta m_{31}^2 (\tan^4{\theta_{12}}-1))^2}{(\Delta m_{21}^2 + \Delta m_{31}^2 (\tan^4{\theta_{12}}-1))^2} \; .
\end{equation}
In the limit of vanishing solar mass splitting we find
\begin{equation}
\label{eq:meem3rel}
|m_{ee}| \approx m_3 \; .
\end{equation}
Taking the best-fit values  $\Delta m_{21}^2=7.65 \cdot 10^{-5} \eV^2$, 
$\Delta m_{31}^2=-2.40 \cdot 10^{-3} \eV^2$ 
and $\sin^2{\theta_{12}}=0.304$ \cite{experimentaldata} we obtain
$s \approx 0.02075$, $t \approx 0.03502$, $x \approx 0.02146$
\footnote{Actually we find four solutions which all lead to the same absolute values, but
to different signs for $s$, $t$ and $x$, with the constraint that $s$ and $x$ have the same sign.} for $v_u \approx 100 \GeV$,
$\Lambda \approx 4 \cdot 10^{11} \GeV$ and $v/\Lambda \approx \lambda^2 \approx 0.04$.
The neutrino masses are $m_1 \approx 0.05348 \, \eV$, $m_2 \approx 0.05419 \, \eV$ 
and $m_3 \approx 0.02146 \, \eV$. Their sum $\sum m_{i} \approx 0.1291 \, \eV$
lies below the upper bound required from cosmological data \cite{cosmobounds}.
$|m_{ee}|$ equals $0.02075 \eV$ which might be detectable in the future \cite{0nubb}. The two Majorana
phases $\phi_{1,2}$ are $\phi_{1}=\pi/2$ and $\phi_{2}=0$.
For tritium $\beta$ decay we find $m_{\beta} \approx 0.05370 \,\eV$ which is about a factor of six smaller
than the expected sensitivity of the KATRIN experiment \cite{katrin}.

\begin{figure}[t]
\parbox[t]{3.5in}{
\epsfig{file=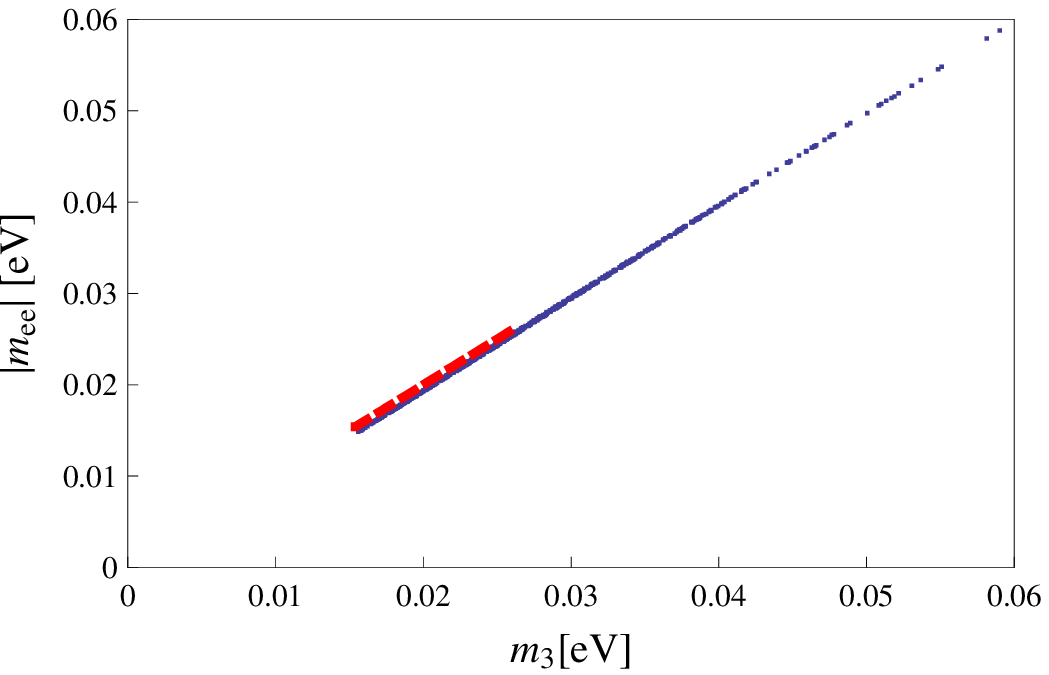,width=7cm,height=5.5cm}
\begin{minipage}[t]{7.5cm}
\caption[]{$|m_{ee}|$ plotted against $m_{3}$ for $z\neq 0$. The dashed (red) line indicates the results for
$z=0$. Mass squared differences and the solar mixing angle are in the allowed $2 \sigma$ ranges
\cite{experimentaldata}. As one can see, $|m_{ee}|$ and $m_3$ have nearly the same value. 
Additionally, one finds that  
$m_3$ has a lower bound around $0.015 \eV$. For $z=0$ we also find an upper bound on $m_3$.
\label{fig:scattera}}
\end{minipage}
}
\parbox[t]{3.5in}{
\epsfig{file=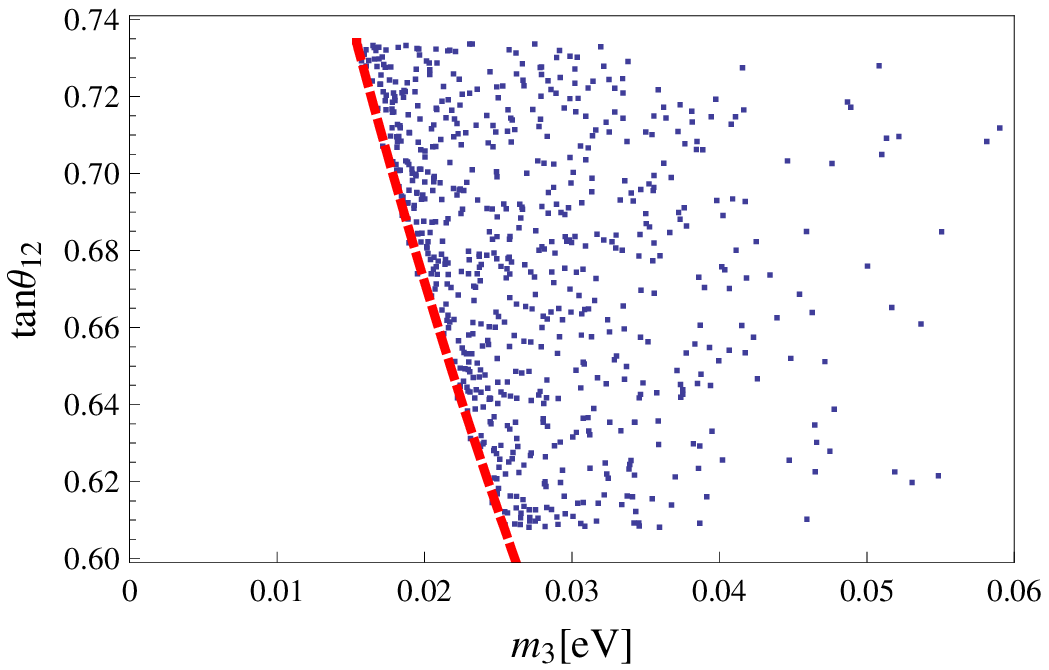,width=7cm,height=5.5cm}
\begin{minipage}[t]{7.5cm}
\caption[]{$\tan{\theta_{12}}$ plotted against $m_3$ for non-vanishing $z$. 
Again the dashed (red) line indicates $z=0$ (assuming the best-fit value for the atmospheric mass squared difference) and gives a lower bound for $z\neq 0$.
Apart from that the results for $\tan \theta_{12}$ are only constrained by the requirement
that they are within the experimental $2 \sigma$ ranges \cite{experimentaldata}, 
$0.61 \lesssim \tan \theta_{12} \lesssim 0.73$. 
\label{fig:scatter2a}}
\end{minipage}
}
\end{figure}
Turning to the general case with $z\neq 0$ we first observe that 
also in this case the light neutrinos have to have an
inverted hierarchy. To see this let us assume that the matrix in \Eqref{eq:Mnu_SCPV} would allow 
the neutrinos to be normally ordered, i.e.
$m_3>m_1$ as well as $m_3>m_2$. From $m_3^2-m_2^2 >0$ then follows
\begin{equation}
x^2+z^2-s^2-4t^2-\sqrt{(s-x)^2 (8t^2+(s+x)^2)+2 (4 t^2+x^2-s^2)z^2+z^4} > 0 \; .
\label{eq:NHineq1}
\end{equation}
From this we can deduce
\begin{equation}
\label{eq:NHineq2pr}
x^2+z^2 > s^2+4 t^2 
\;\;\; \mbox{and} \;\;\;
16 t^2 (t^2+x (s-x) -z^2) > 0 \; .
\end{equation}
Rearranging the first inequality and taking $t\neq 0$ (otherwise $\theta_{12}$ is zero) 
for the second one, we get
\begin{equation}
\label{eq:NHineq2}
x^2-s^2 > 4 t^2-z^2 
\;\;\; \mbox{and} \;\;\;
t^2-z^2 > x (x-s) \; .
\end{equation}
\begin{figure}[t]
\parbox[t]{3.5in}{
\epsfig{file=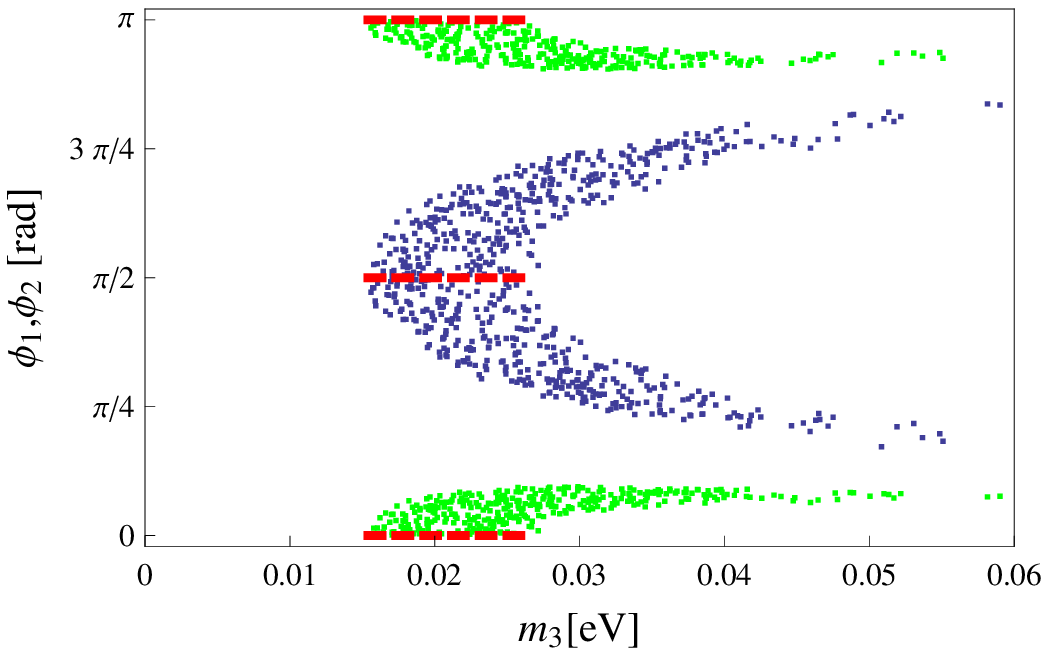,width=7cm,height=5.5cm}
\begin{minipage}[t]{7.5cm}
\caption[]{The Majorana phases $\phi_1$ (blue/darker gray) and $\phi_2$ (green/lighter gray) 
plotted against the lightest neutrino mass $m_3$ for non-vanishing $z$. The values for $z=0$,  
$\phi_1 = \frac{\pi}{2}$, $\phi_2 =0$, are displayed by dashed (red) lines. Notice that the results
for $z\neq 0$ are centered around these values. The measured quantities, $\Delta m_{21}^{2}$,
$|\Delta m_{31}^{2}|$ and $\theta_{12}$, are within the $2 \sigma$ ranges \cite{experimentaldata}.
\label{fig:scatter3a}}
\end{minipage}
}
\parbox[t]{3.5in}{
\epsfig{file=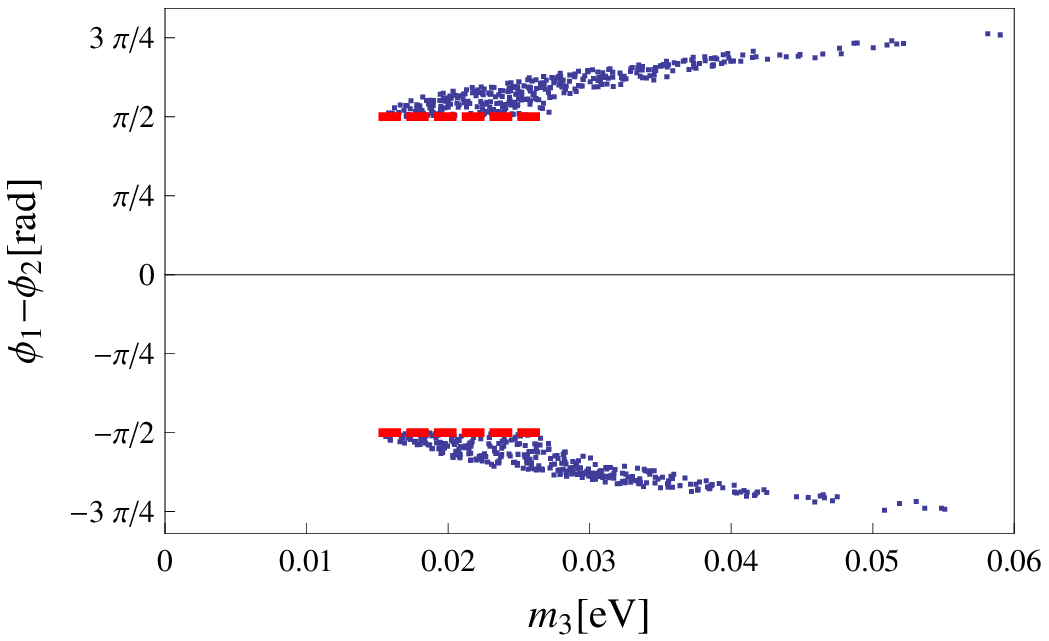,width=7cm,height=5.5cm}
\begin{minipage}[t]{7.5cm}
\caption[]{Phase difference $\phi_1 - \phi_2$ against $m_3$ for $z\neq 0$. 
The case $z=0$, $|\phi_{1}-\phi_{2}|=\pi/2$, 
is given by the dashed (red) lines. As one can see, $|\phi_{1} -\phi_{2}|$ is restricted to
the interval $[\pi/2,3 \pi/4]$ for $m_3 \lesssim 0.06 \eV$. 
Its deviation from $\pi/2$ increases with increasing $m_3$.
Again, the mass squared differences and $\theta_{12}$ are within the experimentally allowed
$2 \sigma$ ranges \cite{experimentaldata}.
\label{fig:scatter4a}}
\end{minipage}
}
\end{figure}
The sum of these inequalities leads to
\begin{equation}
\label{eq:NHineq3}
s (x-s) > 3 t^2 > 0 \; .
\end{equation}
From \Eqref{eq:NHineq3} we see that $s$ and $x$ have the same sign, while $x^2>s^2$, hence $x (x-s) > s (x-s)$. 
Combining \Eqref{eq:NHineq2} and \Eqref{eq:NHineq3}, we find $t^2-z^2 > 3 t^2$, an obvious contradiction. 
Thus, the neutrinos cannot be normally ordered as assumed by $m_3^2>m_2^2$. Instead we always have $m_2^2>m_3^2$
which is only possible in case of an inverted hierarchy. Note that it is a priori not clear that
also $m_1$ is larger than $m_3$, since the size of the mass squared differences has to be tuned so that
$\Delta m_{21}^{2} \ll |\Delta m_{31}^{2}|$.
In fact, $\Delta m_{21}^{2}$ is given by
\begin{equation}
\Delta m_{21}^2 = \Big( \frac{v_u^2}{\Lambda} \Big)^2 \, \Big( \frac{v}{\Lambda} \Big)^{2} \, 
\sqrt{(s-x)^2 (8t^2+(s+x)^2)+2 (4 t^2+x^2-s^2)z^2+z^4} \; .
\end{equation}
It vanishes, if $z=0$ and $s=x$. Thus, $\Delta m_{21}^2 \ll |\Delta m_{31}^2|$ holds, if these equalities
are nearly met. As noted, the vanishing of $z$ can be made a natural result of the model.
The near equality $s \approx x$ however has to be regarded as a certain tuning
of the couplings $y_1$ and $y_4$, see \Eqref{eq:Mnuparameters}. 

We study the general case $z\neq 0$ with a numerical analysis.
To fix the light neutrino mass scale we adjust the resulting solar mass squared difference
to its best-fit value. At the same time the atmospheric mass squared difference and the mixing angle
$\theta_{12}$ have to be within the allowed $2 \sigma$ ranges \cite{experimentaldata}. 
First, we note that our numerical results confirm that $z$ has to be in general smaller than the 
parameters $s$, $t$ and $x$ and that $s$ and $x$ have to have nearly the same value.
In \Figref{fig:scattera} we plotted $|m_{ee}|$ against the lightest neutrino 
mass $m_3$.
As one can see, the approximate equality of $|m_{ee}|$ and $m_3$, deduced for $z=0$
in \Eqref{eq:meem3rel}, still holds for $z \neq 0$. The dashed (red) line is the result for
$z=0$. One finds that $m_3$ has a minimal value around $0.015 \eV$, i.e. $m_3$
cannot vanish, and for $z=0$ it also has a maximal one
around $0.027 \eV$. These two bounds can be found as well by using \Eqref{eq:m3tanth12rel}.
The non-vanishing of $m_3 \approx |m_{ee}|$ agrees with the findings in the literature that $|m_{ee}|$ is required to be
larger than $0.01 \eV$, if neutrinos follow an inverted hierarchy \cite{meebound}.
\Figref{fig:scatter2a} shows that the relation in \Eqref{eq:m3tanth12rel}, which is fulfilled to a good 
accuracy for $z=0$, gives a lower bound for $z\neq0$ in the $\tan \theta_{12}$-$m_3$ plane 
and no further constraints on the solar mixing angle can be derived. Note that we used the best-fit value of the 
atmospheric mass squared difference for the dashed (red) line in \Figref{fig:scatter2a}. Finally, we plot the 
Majorana phases $\phi_1$ and $\phi_{2}$ in \Figref{fig:scatter3a} against the lightest neutrino mass $m_3$.
As one can see, the phase $\phi_1$ (blue/darker gray) varies between $\pi/8$ and $7 \pi/8$, while $\phi_2$ 
(green/lighter gray) either lies in the interval $[0,\pi/8]$ or $[7 \pi/8,\pi]$ for small values of
$m_3$, i.e. $m_3 \lesssim 0.06 \eV$. The dashed (red) lines indicate
again the value of $\phi_{1}$ and $\phi_{2}$ achieved in the limit $z=0$. As the difference 
$\phi_{1}-\phi_{2}$ of the two Majorana phases is the only quantity which can be realistically 
determined by future experiments \cite{0nubb} through
\begin{equation}
|m_{ee}| = |m_1 \cos^2 \theta_{12} e^{2 i \, (\phi_1 - \phi_2)} + m_2 \sin ^2 \theta_{12}| \; ,
\end{equation}
we also plot $\phi_1 - \phi_2$ against $m_3$ in \Figref{fig:scatter4a}. This plot shows that the phase difference 
has to lie in the rather narrow ranges $[-3 \pi/4, -\pi/2]$ or $[\pi/2,3 \pi/4]$ for small values of $m_3$. 
As one can see, the deviations from 
$|\phi_{1}-\phi_{2}|=\pi/2$ ($z=0$ case) become larger for larger values of $m_3$.

\subsection{Flavon Superpotential}
\label{sec:flavons_LO}

In the following we discuss the flavon superpotential and show that the VEV structure assumed in  
(\Eqref{eq:VEVs1} and) \Eqref{eq:VEVs2} naturally arises, 
as does the spontaneous CP violation. In constructing the superpotential we work along the lines of
\cite{A4alignment}. For this purpose, we generalize $R$-parity to a $U(1)_R$ symmetry under which
the ``matter fields'' transform with charge +1, the fields $h_u$ and $h_d$ and the flavons are uncharged and another
type of fields, the driving fields, have charge +2. These fields transform trivially under the 
Standard Model gauge group,
but non-trivially under the flavor symmetry. The set needed for constructing the potential consists of
$\chi_{e}^{0} \sim (\MoreRep{1}{1}, \omega^4)$, 
$\sigma^{0} \sim (\MoreRep{1}{4}, \omega)$ and 
$\chi^{0}_\nu \sim (\MoreRep{1}{1}, \omega)$ under $(D_4,Z_5)$.
Since all terms of the superpotential have to have $U(1)_R$ charge +2, the driving fields cannot couple to the
fermions and can only appear linearly in the flavon superpotential.
The renormalizable $D_4 \times Z_5$ invariant superpotential for flavons and driving fields 
reads
\begin{eqnarray}
\label{eq:wfLO}
w_f = & & a \, \chi^0_e \, \chi_e^2 + b \, \chi^0_e \, \varphi_e^2 \\ \nonumber
  & +&  c  \, \sigma^0 \, (\psi_1^2 - \psi_2^2)  
  + d \, \chi^0_{\nu} \, \psi_1 \, \psi_2  + e \, \chi^0_{\nu} \, \varphi_{\nu}^2 
+ f \, \chi^0_{\nu} \, \chi_{\nu}^2 \; . 
\end{eqnarray} 
Assuming that the flavons acquire their VEVs in the supersymmetric limit we can use the 
F-terms of the driving fields to
determine the vacuum structure of the flavons. The equations
\begin{subequations}
\label{eq:FtermsLO}
\begin{eqnarray}
\frac{\partial w_f}{\partial \chi^0_e}&=& a \, \chi_e^2 + b \, \varphi_e^2 = 0 \; ,\\
\frac{\partial w_f}{\partial \sigma^0}&=& c \, (\psi_1^2 - \psi_2^2) = 0 \; ,\\
\frac{\partial w_f}{\partial \chi^0_{\nu}}&=& d \, \psi_1 \, \psi_2 + e \, \varphi_{\nu}^2 
+ f \, \chi_{\nu}^2 = 0 \; , 
\end{eqnarray}
\end{subequations}
result in 
\begin{equation}
\label{eq:VEVsatLOpr}
\langle \chi_e \rangle = \pm i \, \sqrt{\frac{b}{a}} \, \langle \varphi_e \rangle \; , \;\;
\langle \psi_1 \rangle = \pm \langle \psi_2 \rangle \; , \;\; 
\langle \chi_\nu \rangle  
=\pm i \, \sqrt{\frac{d \, \langle \psi_1 \rangle \,\langle \psi_2 \rangle 
+ e\, \langle \varphi_\nu \rangle^2}
{f}}  
\end{equation}
which can be re-written as 
\begin{equation}
\label{eq:VEVsatLO}
 w_e = \pm i \, \sqrt{\frac{b}{a}} \, u_e \; , \;\;
\langle \psi_1 \rangle = \pm v \; , \;\; 
w =\pm i \, \sqrt{\frac{d \, \langle \psi_1 \rangle \, \langle \psi_2 \rangle + e\, u^2}
{f}}\; . 
\end{equation}
Note that the VEVs $\langle \varphi_e \rangle=u_e$, $\langle \psi_2 \rangle=v$ and 
$\langle \varphi_\nu \rangle=u$ are
unconstrained by the potential. 
Note further that the choice of sign in all cases is independent in
\Eqref{eq:VEVsatLOpr} and \Eqref{eq:VEVsatLO}. For the discussion of
the preserved subgroup structure it is anyway  
only relevant whether $\langle \psi_1 \rangle = \langle \psi_2 \rangle$ or 
$\langle \psi_1 \rangle = -\langle \psi_2 \rangle$. 
For $\langle \psi_1 \rangle = \langle \psi_2 \rangle$ as used in \Eqref{eq:VEVs2}
we conserve a subgroup $Z_2$ of $D_4$ generated by $\rm B$, whereas the relation 
$\langle \psi_1 \rangle = -\langle \psi_2 \rangle$ indicates that the
$Z_2$ subgroup generated by
$\rm B A^2$ is left unbroken. 
This $Z_2$ group is also not a subgroup of the $D_2$ group conserved in the charged lepton sector. 
Thus, the subgroups of the charged lepton and the neutrino sector will be misaligned
in both cases. In this paper we only consider the 
case of $\langle \psi_1 \rangle = \langle \psi_2 \rangle=v$.
\Eqref{eq:VEVsatLO} shows then that the VEVs $w_e$ and $w$ necessarily have to be imaginary, so that
CP is spontaneously violated, if the parameters 
$a, ..., f$ and the VEVs $u_e$, $v$ and $u$ are chosen as positive.

We remark that due to the $U(1)_R$ symmetry a $\mu$-term $\mu h_u h_d$ is forbidden in our model 
and has to be generated by some other mechanism. This feature is shared by all models using a
$U(1)_R$ symmetry. In the derivation of \Eqref{eq:FtermsLO} terms of the form $\chi^0_\nu h_u h_d$ which
couple a driving field to the MSSM Higgs fields
can be safely neglected. They also cannot induce a $\mu$-term, 
since only vanishing VEVs are allowed for the driving fields, if the parameters $a, ..., f$ and the
flavon VEVs are non-zero, as it is in our case.
Finally, note that we find flat directions in this potential in the case of spontaneous CP violation under
discussion here. These are however expected to be lifted by the inclusion of the NLO corrections, see
\Secref{sec:flavons_NLO}, as well as through soft supersymmetry breaking terms. 
Such nearly flat directions might be of interest
for inflation \cite{Antusch:2008gw}.

\section{Next-to-Leading Order Corrections}
\label{sec:NLO}

In order to determine how our results are corrected at NLO, we take into account the effects of operators 
which are suppressed by one more power of the cutoff scale $\Lambda$ compared to the LO. Such
contributions to the fermion masses include two instead of only one flavon. In the flavon superpotential 
we add terms consisting of one driving field and three flavons. It turns out that there are actually no contributions to the fermion 
masses from two-flavon insertions due to the $Z_5$ symmetry. Hence, the only NLO corrections we need to consider are those of the flavon 
superpotential, which lead to a shift in the flavon VEVs parameterized as
\begin{equation}
\label{eq:VEVshifts}
\langle \chi_e \rangle = w_e + \delta w_e
 \; , \;\;
\langle \chi_{\nu} \rangle = w + \delta w  
\;\;\; \mbox{and} \;\;\;
\langle \psi_1 \rangle = v+ \delta v \; .
\end{equation}
The VEVs $\langle \varphi_e \rangle=u_e$, $\langle \varphi_\nu \rangle=u$ and 
$\langle \psi_2 \rangle=v$ which are not determined at
LO remain unconstrained also at NLO. The natural size of the VEV shifts is
\begin{equation}
\frac{\delta \rm VEV}{\rm VEV} \sim \lambda^2 \; .
\end{equation}
As will be discussed in \Secref{sec:flavons_NLO}, the shifts $\delta w$ and $\delta w_e$ are in general
complex, whereas the shift $\delta v$ in the VEV $\langle \psi_1 \rangle$ is real 
for this type of spontaneous CP violation.

\subsection{Fermion Masses}
\label{sec:fermionmasses_NLO}

The VEV shifts induce corrections to the lepton mass matrices given in \Eqref{eq:fermionsatLO} when the shifted 
VEVs are inserted into the LO terms, see \Eqref{eq:wlatLO}. 
In case of the charged lepton masses only the VEV of 
$\chi_e$ is shifted. Such a shift is however not relevant, since it can be absorbed into the
Yukawa couplings $y^e_1$ and $y^e_2$. \footnote{These then become complex which however does not affect our 
results.} Especially, $U_l$ is still given by \Eqref{eq:Ul}.
The form of the neutrino mass matrix is changed through the shifts of the VEVs into
\begin{equation}
\label{eq:MnuNLO}
M_{\nu} = \frac{v^2_u}{\Lambda^2} 
 \left( 
 \begin{array}{ccc}  
 y_1 (w + \delta w) & y_2 v & y_2 (v+\delta v) \\ 
 y_2 v & y_3 u &  y_4 (w + \delta w) \\ 
 y_2 (v+ \delta v) &  y_4 (w + \delta w) & y_3 u 
 \end{array} 
 \right) \; . 
\end{equation}
Note that $\delta w$ cannot be simply absorbed into $w$, since $\delta w$ is complex, whereas $w$ is imaginary. 
In the charged lepton mass basis the matrix in 
\Eqref{eq:MnuNLO} reads
\begin{equation}
\label{eq:MnuprNLOpr}
M_{\nu}' = \frac{v^2_u}{\Lambda^2} 
\left( 
\begin{array}{ccc}  
y_1 (w + \delta w) & y_2 (v+ e^{i \pi/4}\delta v /\sqrt{2}) & y_2 (v+ e^{-i \pi/4}\delta v/\sqrt{2}) \\ 
y_2 (v+ e^{i \pi/4}\delta v/\sqrt{2}) & y_4 (w + \delta w) & y_3 u  \\ 
y_2 (v+ e^{-i \pi/4}\delta v/\sqrt{2}) &  y_3 u & y_4 (w + \delta w) 
\end{array} 
\right) \; .
\end{equation}
To evaluate the shifts in the neutrino masses and to discuss the deviations of the mixing angles
from their LO values, especially $\theta_{13}$ from zero and $\theta_{23}$ from maximal, we parameterize 
the Majorana neutrino mass matrix as
\begin{equation}
\label{eq:MnuprNLO}
M_{\nu}' = \frac{v_u^2}{\Lambda} \, \frac{v}{\Lambda}
\left( 
\begin{array}{ccc}  
  i \, s \, (1 + \alpha \, \epsilon) & t \, (1 + e^{i \pi/4} \, \epsilon) &  t \, (1 + e^{-i \pi/4} \, \epsilon)\\ 
  t \, (1 + e^{i \pi/4} \, \epsilon) &  i \, x \, (1 + \alpha \, \epsilon) &  z\\ 
  t \, (1 + e^{-i \pi/4} \, \epsilon) &  z & i \, x \, (1 + \alpha \, \epsilon) 
\end{array} 
\right)
\end{equation}
with $s$, $t$, $x$ and $z$ as given in \Eqref{eq:Mnuparameters} and
\footnote{We assume that $\epsilon$ is positive.}
\begin{equation}
\alpha \, \epsilon = \frac{\delta w}{w} \; , \;\; \alpha = \alpha_r + i \, \alpha_i
\;\;\; \mbox{and} \;\;\; 
\epsilon = \frac{1}{\sqrt{2}} \, \frac{\delta v}{v} \approx \lambda^2 \approx 0.04 \; .
\end{equation}
The neutrino masses and mixing parameters resulting from \Eqref{eq:MnuprNLO} 
can then be calculated in an expansion in the small parameter $\epsilon$. 
We observe that the mass shift of $m_3^2$ would vanish for $\delta w$ being zero.
Its explicit form is
\begin{equation}
(m_3^{\rm NLO})^2 = (m_3^{\rm LO})^2 + 2 \, \Big( \frac{v_u^2}{\Lambda} \Big)^2 \, \Big( \frac{v}{\Lambda} \Big)^2 \, 
x (\alpha_r \, x +\alpha_i \, z) \, \epsilon
\end{equation}
with $(m_3^{\rm LO})^2$ given in \Eqref{eq:numasses}.
Similarly, the masses $m_1^2$ and $m_2^2$ undergo shifts proportional to $\epsilon$. A simple
expression can however only be found for the sum $m_1^2 + m_2^2$
\begin{equation}
(m_1^{\rm NLO})^2 + (m_2^{\rm NLO})^2 = (m_1^{\rm LO})^2 + (m_2^{\rm LO})^2 + 2 \, \Big( \frac{v_u^2}{\Lambda} \Big)^2 \, \Big( \frac{v}{\Lambda} \Big)^2 \, (2 \, \sqrt{2} \, t^2+\alpha_r \, (s^2+x^2)-\alpha_i \, x \, z) \, \epsilon
\end{equation}
$(m_{1,2}^{\rm LO})^2$ can be found in \Eqref{eq:numasses}.
The mixing angle $\theta_{13}$ no longer vanishes and we find
\begin{equation}
\label{eq:sintheta13}
\sin \theta_{13} \approx \left|\frac{t \, x}{t^2+(s-x)x-z^2}  \right| \, \epsilon \; .
\end{equation}
For $\theta_{23}$ we get
\begin{equation}
\tan \theta_{23} \approx 1 + \sqrt{2} \, \frac{x \, z}{t^2+(s-x)x-z^2} \, \epsilon \; .
\end{equation}
The deviation from maximal atmospheric mixing can also be expressed through
 \begin{equation}
\label{eq:cos2theta23}
|\cos 2 \theta_{23}| 
\approx  \sqrt{2} \, \left| \frac{x \, z}{t^2+(s-x)x-z^2}  \right| \, \epsilon 
\approx \sqrt{2} \, \left| \frac{z}{t} \right| \, \sin \theta_{13} \; .
\end{equation}
From both formulae one can deduce that in the case $z=0$ the corrections to maximal atmospheric mixing
are not of the order $\epsilon$, but only arise at $\mathcal{O}(\epsilon^2)$. Contrary to this
$\theta_{13}$ still receives corrections of order $\epsilon$, if $z=0$.
The solar mixing angle $\theta_{12}$ which is not fixed to a precise value in this model
also gets corrections of order $\epsilon$. We note that the
smallness of $|s-x|$ and $z$, required by the smallness of $\Delta
m_{21}^2$, might lead to a disturbance of the expansion in the
parameter $\epsilon$.

A correlation between $\cos 2 \theta_{23}$ and $\sin \theta_{13}$ depending only on physical
quantities, $\Delta m_{ij}^{2}$, ..., and not on the parameters of the model, $s$, $t$, ...,
can be obtained by an analytic consideration which is done analogously to the study performed
in \cite{GLS3}. Clearly, the matrix in \Eqref{eq:MnuprNLOpr} is no longer $\mu-\tau$ symmetric,
however we find the following remnants of this symmetry
\begin{equation}
\label{eq:mutaurest}
(M_{\nu}')_{e \mu} = (M_{\nu}')_{e \tau}^{\ast}  \;\;\; \mbox{and} \;\;\;
(M_{\nu}')_{\mu \mu} = (M_{\nu}')_{\tau \tau} \; .
\end{equation}
\Eqref{eq:mutaurest} shows that $\mu-\tau$ symmetry is only broken
by phases, but not by the absolute values of the matrix elements. This
leads to
\begin{eqnarray}
&& 0=|(M_{\nu}')_{e \mu}|^2+|(M_{\nu}')_{\mu \mu}|^2-|(M_{\nu}')_{e \tau}|^2-|(M_{\nu}')_{\tau \tau}|^2 \\ 
&& 0=(M_{\nu}' M_{\nu}^{\prime \, \dagger})_{\mu \mu} - (M_{\nu}' M_{\nu}^{\prime \, \dagger})_{\tau \tau}
= \sum_{j=1}^{3} m_j^2(|(U_{MNS})_{\mu j}|^2-|(U_{MNS})_{\tau j}|^2 )\nonumber \\ \nonumber
&& 0=  \Big((\sin ^2 \theta_{12}-\sin^2 \theta_{13} \cos^2 \theta_{12}) \, m_1^2 
+ (\cos^2 \theta_{12}-\sin^2 \theta_{13} \sin^2 \theta_{12}) \, m_2^2 - \cos^2 \theta_{13} \, m_3^2\Big) 
\cos (2 \theta_{23})\\
&& \phantom{0= \;\;\;\;}
 -  \Delta m_{21}^2 \sin 2 \theta_{12} \sin 2 \theta_{23} \cos \delta \sin \theta_{13}  \; .
\end{eqnarray}
Since $\sin \theta_{13} \sim \mathcal{O}(\epsilon)$ and $\cos 2 \theta_{23} \sim \mathcal{O}(\epsilon)$
is already known, we can linearize this equation and obtain
(using best-fit values for the physical quantities and the fact that neutrinos have an inverted hierarchy in this model)
\begin{equation}
\label{eq:cos2th23s13rel}
\cos 2 \theta_{23} \approx -\frac{\Delta m_{21}^2 \sin 2 \theta_{12}}{\Delta m_{32}^2 + \Delta m_{21}^2 \sin^2 \theta_{12}} \cos \delta \sin \theta_{13} \approx 0.03  \cos \delta \sin \theta_{13} \; .
\end{equation}
\Eqref{eq:cos2th23s13rel} can be used to estimate the largest possible deviation from maximal mixing. 
For $\sin \theta_{13}$ being at its $2 \, \sigma$ limit of $0.2$ and $|\cos \delta|=1$, $|\cos 2 \theta_{23}|$ still has to be less than $6 \times 10^{-3}$ which is well within the $1 \, \sigma$ error. 
Finally, we note that \Eqref{eq:cos2th23s13rel} must be consistent with \Eqref{eq:cos2theta23} and thus we again find that $z$ ought to be small.

\subsection{Flavon Superpotential}
\label{sec:flavons_NLO}

The corrections to the flavon superpotential stem from terms involving one driving field and three flavons.
These terms are non-renormalizable and suppressed by the cutoff scale $\Lambda$. 
We find
\begin{eqnarray}
\Delta w_{f}= & & \frac{k_1}{\Lambda} \, \chi_e^0 \, \chi_{\nu}^3 
+ \frac{k_2}{\Lambda} \, \chi_e^0 \,\chi_{\nu} \, \varphi_{\nu}^2 
+ \frac{k_3}{\Lambda} \, \chi_e^0 \, \chi_{\nu} \psi_1 \psi_2 
+ \frac{k_4}{\Lambda} \, \chi_e^0 \, \varphi_{\nu} \, (\psi_1^2 + \psi_2^2)\\ \nonumber
&+& \frac{k_5}{\Lambda} \, \sigma^0 \, \varphi_e \, \chi_e^2 
+ \frac{k_6}{\Lambda} \, \sigma^0 \, \varphi_e^3 
+ \frac{k_7}{\Lambda} \,  \chi_{\nu}^0 \, \chi_e^3 
+ \frac{k_8}{\Lambda} \,  \chi_{\nu}^0 \, \chi_e \varphi_e^2 \, .
\end{eqnarray}
Assuming that CP is only spontaneously violated forces all $k_i$ to be real.
We calculate the F-terms of $w_f + \Delta w_f$ for the driving fields using that the VEVs 
can be parameterized as 
\begin{equation}
\langle \chi_e \rangle = w_e + \delta w_e
 \; , \;\;
\langle \chi_{\nu} \rangle = w + \delta w  
\;\;\; \mbox{and} \;\;\;
\langle \psi_1 \rangle = v+ \delta v \; .
\end{equation}
The VEVs $\langle \varphi_e \rangle=u_e$, $\langle \varphi_\nu \rangle=u$ and 
$\langle \psi_2 \rangle=v$ are not determined at LO.
We assume that only terms
containing up to one VEV shift or the suppression factor $1/\Lambda$, but not both, are relevant.
The F-terms then lead to
\begin{subequations}
\label{eq:determineVEVshifts}
\begin{eqnarray}
&&2 \, a \, w_e \, \delta w_e + \frac{1}{\Lambda} (k_1 \, w^3 + k_2 \, u^2 \, w 
+ k_3 \, v^2\, w+2 \, k_4 \, u \, v^2) = 0 \; ,\\
&&2 \, c \, v \, \delta v+ \frac{u_e}{\Lambda} (k_5 \, w_e^2+k_6 \, u_e^2) = 0 \; ,\\
&&d \, v \, \delta v+2 \, f \, w \, \delta w + \frac{w_e}{\Lambda} (k_7 \, w_e^2+k_8 \, u_e^2) = 0 \; .
\end{eqnarray}
\end{subequations}
Here we have chosen the solutions with $+$ in \Eqref{eq:VEVsatLO}. 
The explicit form of the shifts reads
\begin{subequations}
\label{VeVshift2}
\begin{eqnarray}
&&\delta v  = -\frac{1}{2 \, c} \, \frac{u_e}{v \, \Lambda} \, (k_5 \, w_e^2 + k_6 \, u_e^2) \; , \\
&&\delta w = \frac{1}{4 \, c \, f} \, \frac{1}{w \, \Lambda} \, (d \, (k_5 w_e^2 + k_6 u_e^2) \, u_e - 2 \, c 
\, (k_7 \, w_e^2 + k_8 \, u_e^2) \, w_e) \; , \\
&&\delta w_e = -\frac{1}{2 \, a} \, \frac{1}{w_e \, \Lambda} \, (k_1 \, w^3 + k_2 \, u^2 \, w
+ k_3 \, v^2 \, w + 2 \, k_4 \, u \, v^2) \; .
\end{eqnarray}
\end{subequations} 
As one can see, for our type of spontaneous CP violation $\delta v$ is real, whereas
$\delta w_e$ and $\delta w$ turn out to be complex in general. 
As can be read off from \Eqref{VeVshift2} all shifts are generically of order
\begin{equation}
\frac{\delta \rm VEV}{\rm VEV} \sim \lambda^2 \;\;\; \mbox{for} \;\;\; \rm VEV \sim \lambda^2 \, 
\Lambda \; .
\end{equation}
Finally, note that the free parameters $\langle \varphi_e \rangle=u_e$, $\langle \varphi_\nu \rangle=u$ 
and $\langle \psi_2 \rangle=v$ are still undetermined.

\section{Summary and Outlook}
\label{sec:summary}

We constructed a supersymmeterized version of the $D_4$ model by Grimus and Lavoura \cite{GL1}.
For this purpose, we replaced the Higgs doublets transforming under the flavor group $D_4$ by
gauge singlets. We also enlarged the auxiliary symmetry which separates the different
flavor breaking sectors from $Z_2^{(aux)}$ to $Z_5$. The simplest supersymmeterized $D_4$
model does not contain right-handed neutrinos, but neutrinos get masses through the
operator $l h_u l h_u/\Lambda$. Apart from these slight changes the model is essentially
the same as the one by GL, since we also generate maximal atmospheric mixing
and vanishing $\theta_{13}$ through the fact that $D_4$ is broken to $D_2$ in
the charged lepton and to $Z_2$ in the
neutrino sector. The crucial issue of the vacuum alignment 
is elegantly solved through an appropriately constructed flavon potential.
We performed a phenomenological analysis under the assumption of a certain type of spontaneous CP violation
suggested by the minimization of the potential.
As a result, the neutrinos have to have an inverted hierarchy. The quantity $|m_{ee}|$, measured in $0\nu\beta\beta$
decay, is almost equal to the lightest neutrino mass $m_3$. Furthermore, we found
that $m_3$ cannot vanish and has a lower bound around $0.015 \eV$. The Majorana phases $\phi_{1}$
and $\phi_{2}$ are restricted to a certain range at least for small $m_3$. In contrast to that the solar mixing angle
$\theta_{12}$ can take all values allowed by experiments. We also analyzed the NLO
terms in this model and showed that they only induce shifts in the VEVs of the flavons, but
no additional terms in the Yukawa sector appear.
The shifts yield deviations from the LO results, $\theta_{13}=0$ and $\theta_{23}=\pi/4$.
Comparing these deviations we see that although both of them could in principle be
of order $\epsilon \approx \lambda^{2} \approx 0.04$, the smallness of the parameter $z$, necessary to
arrive at mass squared differences and $\theta_{12}$  within the $2\sigma$ ranges, leads to
the fact that $\theta_{23}$ is much closer to $\pi/4$ than $\theta_{13}$ to zero.

The supersymmeterization of the $D_4$ model has to be regarded as a step towards a
grand unified model with a dihedral flavor symmetry for two reasons: (a) low scale supersymmetry
elegantly cures the hierarchy problem and easily allows the gauge
couplings to be unified at $10^{16} \GeV$ and (b) the replacement of the Higgs doublets 
transforming under the flavor group by flavons is important for disentangling the
breaking of flavor and gauge groups. However, another essential feature of a unified theory is
 that the lepton sector cannot be discussed without also considering quarks.
So, one of the major challenges to tackle is the question how to implement the quark masses
and their mixings in a model with a dihedral symmetry. In the recent past models
have been presented which are able to predict the Cabibbo angle with the help of the 
flavor group $D_7$ and $D_{14}$ \cite{dntheory,thetaC,symmetrygeneral}. 
Thus, it is interesting to search for a way to combine these
models and to find a (probably larger) dihedral symmetry leading to the same results, which we get from a $D_4$ flavor group in the lepton sector
and from a $D_7$ or $D_{14}$ group in the quark sector.

\subsection*{Acknowledgements}

\noindent We would like to thank W. Rodejohann and M. A. Schmidt for discussions.
A.B. acknowledges support from the Studienstiftung des Deutschen Volkes.
C.H. was supported by the ``Sonderforschungsbereich'' TR27.

\appendix

\section{Connection to the Group Basis chosen in \cite{GL1}}
\label{appA}

Note that replacing the left-handed fields $l$ by $U_l^{\star} \, l$, with $U_l$ given in \Eqref{eq:Ul}, 
is equivalent to changing the basis in which
the generators $\rm A$ and $\rm B$ are given for the two-dimensional representation of $D_4$. 
Since also the second and third generation of the right-handed charged leptons form a doublet under 
$D_4$, we also have to transform them to show that this corresponds to a change of the generator basis of the
$D_4$ doublet. By calculating the matrix $U_{e^c}$ which diagonalizes
$M_{l}^{\dagger} M_{l}$ one finds that $U_{e^c}$ equals $U_{l}^{\star} \, P$ where
$P$ is a diagonal phase matrix. This matrix $P$ induces an unphysical rephasing of the right-handed fields 
to keep $U^{\dagger}_l \, M_l \, U_{e^c}$ a diagonal matrix with positive entries. The change of basis 
(induced by the unitary matrix $U_l^{\star}$) leads to real generators $\rm A$ and $\rm B$ of the form
\begin{equation}
\rm A = \left(
\begin{array}{cc}
	0 	& -1\\
	1	& 0
\end{array}
\right) \;\;\; \mbox{and} \;\;\;
\rm B = \left(
\begin{array}{cc}
	0 	& 1\\
	1	& 0
\end{array}
\right) \; .
\end{equation}
This coincides with the basis chosen in \cite{GL1}, if $\rm A$ is identified with the 
product $h g$ and $\rm B$ with the generator $h$. The identifications  
for the singlets are the following: $\MoreRep{1}{$++$}$ corresponds to $\MoreRep{1}{1}$,
$\MoreRep{1}{$+-$}$ to $\MoreRep{1}{4}$, $\MoreRep{1}{$-+$}$ to $\MoreRep{1}{3}$
and $\MoreRep{1}{$--$}$ to $\MoreRep{1}{2}$. 
In \cite{GL1} the charged lepton
mass matrix is diagonal without any further transformation and $\theta_{13}=0$ and $\theta_{23}$ maximal 
can be directly read off from the neutrino mass matrix.
This is the same in our case, if we go into the primed basis, see $M_\nu^\prime$ in \Eqref{eq:Mnu}.

\section{Importance of Mismatch of Subgroups}
\label{appB}

To elucidate the reason why the two subgroups preserved in the charged lepton and the neutrino
sector have to be different, i.e. the $Z_2$ subgroup present in the neutrino sector should not be a
subgroup of the $D_2$ group of the charged lepton sector, observe that $M_{l} \, M_{l}^{\dagger}$ 
as well as $M_{\nu} \, M_{\nu}^{\dagger}$ for $M_l$ and $M_{\nu}$ given in \Eqref{eq:fermionsatLO}
can be written in the following form
\begin{equation}
\label{eq:MiMidagger}
M_i \, M_i ^{\dagger} =
\left(
\begin{array}{ccc}
	A_i & B_i \, e^{i \, \beta_i} & B_i \, e^{i \, (\beta_i + \phi_i \, \mathrm{j})}\\
	B_i \, e^{- i \, \beta_i} & C_i & D_i \, e^{i \phi_i \, \mathrm{j}}\\
	B_i \, e^{-i \, (\beta_i + \phi_i \, \mathrm{j})} & D_i \, e^{-i \, \phi_i \, \mathrm{j}} & C_i
\end{array}
\right) \;\;\;\;\;\;\; i=l,\nu \; .
\end{equation}
This form is achieved for $M_l$ ($M_\nu$) as long as at least a $Z_2$ group, originating from 
$\mathrm{B} \, \mathrm{A}^{m}$, is conserved in the charged lepton (neutrino) sector.
A matrix of this type is diagonalized through
\begin{equation}
\label{eq:Ugeneral}
U_i = 
\left(
\begin{array}{ccc}
	e^{i \, \beta_i} & 0 & 0\\
	0 & 1 & 0\\
	0 & 0 & e^{-i \, \phi_i \, \mathrm{j}}
\end{array}
\right) \, U_{\mathrm{max}} \, U_{12} (\theta_i) \, U (\alpha_k^i)
\end{equation}
where
\begin{equation}
U_{\mathrm{max}} = \left( 
\begin{array}{ccc}
	1 	& 0 			& 0\\
 	0 	& \frac{1}{\sqrt{2}}	& -\frac{1}{\sqrt{2}}\\
 	0 	& \frac{1}{\sqrt{2}}	& \frac{1}{\sqrt{2}}
\end{array}
\right) \; , \;\;
U_{12} (\theta_i) = \left(
\begin{array}{ccc}
	\cos \theta_i & \sin \theta_i & 0\\
	-\sin \theta_i & \cos \theta_i & 0\\
	0 & 0 & 1
\end{array}
\right) \; , \;\; U (\alpha_k^i) = \left(
\begin{array}{ccc}
	e^{i \, \alpha_1 ^i} & 0 & 0\\
	0 & e^{i \, \alpha_2 ^i} & 0\\
	0 & 0 & e^{i \, \alpha_3 ^i}
\end{array}
\right) \; .
\end{equation}
In \cite{dntheory,thetaC} it has been shown that the quantities $\phi_i$ and $\mathrm j$ are related to the group
theoretical indices of the flavor symmetry. $\mathrm j$ is the representation index of the 
doublet under which two of the three left-handed lepton generations transform. Thus, it is the 
same for charged leptons and neutrinos. $\phi_i$ can be expressed as
\begin{equation}
\label{eq:groupphase}
\phi_i = \frac{2 \, \pi}{n} \, m_i
\end{equation}
where $n$ is the index of the group $D_n$ and $m_{l \, (\nu)}$ the index of the preserved subgroup in the
charged lepton (neutrino) sector that has a generator of the form $\mathrm{B} \, \mathrm{A}^{m_{l \, (\nu)}}$.
$m_{l \, (\nu)}$ is an integer number between zero and $n-1$.
The parameters $A_i$, ..., $D_i$ and the phase $\beta_i$ are real functions of the matrix entries
of $M_i \, M_i^{\dagger}$, whose proper form is not needed here. The phases $\alpha_k^i$ are irrelevant
for the diagonalization of $M_i \, M_i^{\dagger}$, but are necessary for the diagonalization of the neutrino
mass matrix $M_{\nu}$ alone. The angle $\theta_i$ can be expressed through the parameters $A_i$, ..., $D_i$
as follows
\begin{equation}
\label{eq:tanform}
\tan 2 \theta_i = \frac{2 \sqrt{2} \, B_i}{C_i + D_i -A_i} \; .
\end{equation}
The general form of the MNS matrix is then
\begin{equation}
\label{eq:UMNSgeneral}
U_{MNS} = U^{T}_l \, U_{\nu}^{\star} 
= U (\alpha_k^l) \, U_{12}^{T} (\theta_l) \, U_{\mathrm{max}}^{T} \, 
	\left( 
	\begin{array}{ccc}
		e^{i \, (\beta_l - \beta_\nu)} & 0 & 0\\			
		0 & 1 & 0\\
		0 & 0 & e^{-i \, (\phi_l - \phi_\nu) \, \mathrm{j}}
	\end{array}
	\right)
\, U_{\mathrm{max}} \, U_{12} (\theta_\nu) \, U (-\alpha_{\tilde{k}} ^\nu) \; .
\end{equation}
This form already shows that it is essential to have a non-trivial
phase $e^{-i \, (\phi_l - \phi_\nu) \, \mathrm{j}}$ in order to
guarantee that the maximal mixing in the $2-3$ sector is not cancelled.
For the third column of $U_{MNS}$, which determines 
the mixing angles $\theta_{13}$ and $\theta_{23}$, we find
\begin{eqnarray}\nonumber
&&|(U_{MNS})_{e 3}|= |\sin ((\phi_l -\phi_\nu) \, \mathrm{j}/2) \, \sin \theta_l |\; , \;\;
|(U_{MNS})_{\mu 3}|= |\sin ((\phi_l -\phi_\nu)  \, \mathrm{j}/2) \, \cos \theta_l |\; , \;\;\\
&&|(U_{MNS})_{\tau 3}|= |\cos ((\phi_l -\phi_\nu)  \, \mathrm{j}/2) |\; . \;\;
\label{eq:UMNS3rdrow}
\end{eqnarray}
Using that we preserve a $Z_2$ symmetry generated by $\rm B$ in the neutrino sector
and a $D_2$ group generated by $\mathrm{B} \, \mathrm{A}$ (according to our
convention for the generators of the group $D_2$ introduced in \Secref{sec:grouptheory}) 
in the charged lepton sector, gives for 
$\phi_\nu$ and $\phi_l$
\begin{equation}
\phi_\nu =0 \;\;\; \mbox{and} \;\;\; \phi_l = \frac{\pi}{2} \; .
\end{equation}
$\rm j$ is trivially one, since $D_4$ only contains one irreducible two-dimensional
representation $\Rep{2}$.
As the elements $(1,k)$ and $(k,1)$ with $k=2,3$ in $M_l$ vanish, see \Eqref{eq:fermionsatLO},
the parameter $B_l$ in \Eqref{eq:MiMidagger}
is zero (and also $\beta_l=0$) and thus $\theta_l=0$ as well according to \Eqref{eq:tanform}.
This results in
\begin{equation}
|(U_{MNS})_{e 3}|= 0 \; , \;\;
|(U_{MNS})_{\mu 3}|= |(U_{MNS})_{\tau 3}|= \frac{1}{\sqrt{2}} 
\end{equation}
giving maximal atmospheric mixing and vanishing $\theta_{13}$. A few things are interesting to 
notice: In principle four different cases might occur. These arise from whether 
the subgroups $D_2$ and $Z_2$ contain the same element $\mathrm{B} \mathrm{A}^{m}$ 
or not and from whether the $D_2$ subgroup is unbroken in the charged lepton sector
or only a $Z_2$ subgroup is preserved. The first issue determines whether $m_l$ equals
$m_\nu$ or not, i.e. whether $|\phi_l-\phi_\nu|$ is zero or not. 
The second one is responsible for (non-)zero $\theta_l$. We can see from \Eqref{eq:UMNS3rdrow}
that for no mismatch of the subgroups $\theta_{13}$ as well as $\theta_{23}$ vanish,
in contrast to what is observed in nature. So the mismatch of the two subgroups is necessary.
If $\theta_l$ is zero, i.e. the subgroup present in the charged lepton sector is a $D_2$
group, $\theta_{13}=0$ and $\theta_{23}$ maximal follow. If however only a smaller $Z_2$
group is present in the charged lepton sector, neither $\theta_{13}$ being zero nor
$\theta_{23}$ being maximal holds. Then only the MNS matrix element $|(U_{MNS})_{\tau 3}|$
is fixed by group theory.

Finally, the matrix $U_l$ given in \Eqref{eq:Ul} equals the matrix shown in \Eqref{eq:Ugeneral},
if we additionally set the phases to $\alpha_1^l=0$, $\alpha_2^l=\pi/4$ and $\alpha_3^l=3 \pi/4$.

One might ask the question what actually determines the size of the solar mixing angle $\theta_{12}$
in this context. 
For $\theta_l=0$ we find from \Eqref{eq:UMNSgeneral} that
\begin{equation}
|(U_{MNS})_{e 1}|= |\cos \theta_\nu| \;\;\; \mbox{and} \;\;\; |(U_{MNS})_{e 2}|= |\sin \theta_\nu|
\end{equation}
which shows that $\theta_{12}$ is given by $\theta_\nu$. Since this angle would vanish, if a $D_2$
group instead of a $Z_2$ group (with generator $\mathrm{B} \mathrm{A}^{m}$) was present in the neutrino sector, 
one might interpret the size
of the solar mixing angle as hint to how strongly a $D_2$ group is broken in the neutrino sector.

\end{document}